\documentclass[aps,prl,twocolumn,superscriptaddress,showpacs]{revtex4-1}
\usepackage[colorlinks, linkcolor=blue, anchorcolor=blue, citecolor=blue]{hyperref}
\usepackage{amsmath}
\usepackage{graphicx}
\usepackage{braket}
\usepackage{color}
\usepackage{soul}

\begin{document}

\title{Non-collinear magnetic structure and multipolar order in Eu$_2$Ir$_2$O$_7$}

\author{Yilin Wang}
\affiliation{ Beijing National Laboratory for Condensed Matter Physics, 
              and Institute of Physics, 
              Chinese Academy of Sciences, 
              Beijing 100190, 
              China }

\author{Hongming Weng}
\affiliation{ Beijing National Laboratory for Condensed Matter Physics, 
              and Institute of Physics, 
              Chinese Academy of Sciences, 
              Beijing 100190, 
              China }
              
\author{Liang Fu}              
\affiliation{ Department of physics, Massachusetts Institute of Technology, Cambridge, MA 02139, USA }
           
\author{Xi Dai}  
\affiliation{ Beijing National Laboratory for Condensed Matter Physics, 
              and Institute of Physics, 
              Chinese Academy of Sciences, 
              Beijing 100190, 
              China }
\affiliation{Collaborative Innovation Center of Quantum Matter, 
             Beijing 100190, 
             China}

\date{\today}

\begin{abstract}
The magnetic properties of the pyrochlore iridate material Eu$_2$Ir$_2$O$_7$ (5$d^5$) have been studied based on the first principle calculations, 
where the crystal field splitting $\Delta$, spin-orbit coupling (SOC) $\lambda$ and Coulomb interaction $U$ within Ir 5$d$ orbitals 
are all playing significant roles. The ground state phase diagram has been obtained with respect to the strength of SOC and Coulomb interaction $U$, 
where a stable anti-ferromagnetic ground state with all-in/all-out (AIAO) spin structure has been found. Besides, another anti-ferromagnetic 
states with close energy to AIAO have also been found to be stable. The calculated nonlinear magnetization of the two stable states 
both have the $d$-wave pattern but with a $\pi/4$ phase difference, which can perfectly explain the experimentally observed 
nonlinear magnetization pattern. 
Compared with the results of the non-distorted structure, it turns out that the trigonal lattice distortion is crucial for stabilizing the AIAO state 
in Eu$_2$Ir$_2$O$_7$. Furthermore, besides large dipolar moments, we also find considerable octupolar moments in the magnetic states.

\end{abstract}

\maketitle

The ordering of electronic states is one of the fundamental problems in condensed matter physics. In $3d$ transition metal compounds,  
the ordered states can be described quite well by the product of orders in orbital and spin subspaces~\cite{imada:1998}, 
because the spin-orbit coupling (SOC) here is weak and can be treated perturbatively. While in rare-earth compounds~\cite{santini:2009}, 
the SOC is strong enough to bind the orbital and spin degrees of freedom into rigid objects described by the total angular momentum, and 
the ordered states can then be well understood in terms of the moments with high angular momentum which splits into atomic multiplets under 
crystal field. Unlike the above two limits, the situation in $4d$ and $5d$ transition metal compounds is unique~\cite{william:2014}. 
On one hand, the SOC is strong enough to combine the orbital and spin degrees of freedom to form some complex orders. While on the other hand, 
the SOC is still far away from the limit where the low energy physics can be entirely determined within the subspace with a fixed total angular 
momentum. In fact, the SOC strength in these compounds is comparable with that of the crystal field so that the magnetic orders there can 
involve multiple total angular momentum states. 

The pyrochlore iridates~\cite{nobuyuki:2001,gardner:2010} are typical $5d$ transition metal compounds with many novel properties already 
being discussed extensively in the literatures, including the non-collinear magnetic order~\cite{wan:2011,ybkim:2012,disseler:2012,witczak:2013,
zhao:2011,shinaoka:2015,prando:2016}, the metal-insulator transition~\cite{daiki:2001,kazuyuki:2007,ybkim:2010,kazuyuki:2011,ishikawa:2012,
tafti:2012,sakata:2011,hongbin:2017}, anomalous Hall effect~\cite{machida:2007,machida:2010, xianghu:2012}, topological insulator and Weyl 
semimetal phase~\cite{wan:2011,pesin:2010,ybkim:2010,ybkim:2012,william:2014,sushkov:2015,chen:2012,kargarian:2011,moyuru:2011,guo:2009,chen:2015, 
xianghu:2015}, and the chiral metallic states in the domain wall~\cite{yamaji:2014,ma:2015,yamaji:2016}. Among them, the magnetic order is 
the most fundamental one which determines most of the electronic properties. In Ref.~\cite{wan:2011}, by using the density functional theory plus $U$ (DFT+$U$) 
method Wan \textit{et al.} 
obtained an all-in/all-out (AIAO) order in Y$_2$Ir$_2$O$_7$, which will generate the Weyl semimetal phase if the value of the order parameter 
falls into a proper region. Most recently, Liang \textit{et al.}~\cite{liang:2017} have systematically studied the nonlinear 
magnetization in Eu$_2$Ir$_2$O$_7$ ($5d^5$) by using torque magnetometry, where a magnetic field is applied in $a$-$b$ plane and continuously 
rotated around $c$-axis by $2\pi$. They found a nonlinear magnetization normal to the $a$-$b$ plane with a $d$-wave pattern as a function of 
the rotation angle. Surprisingly, they also found that the $d$-wave pattern has a $\pi/4$ phase shift when the direction of the field 
$\boldsymbol{H_{\text{fc}}}$ applied during the field cooling process changes from $[\bar{1}\,\bar{1}\,0]$ to $[\bar{1}10]$. Their results 
indicate that there is another low energy metastable magnetic structure besides AIAO, which might be stabilized by the field cooling processes. 

Inspired by the experiments on the nonlinear magnetization in Eu$_2$Ir$_2$O$_7$~\cite{liang:2017}, in the present letter, 
we reexamine the magnetic structure of the pyrochlore iridates. By using the DFT together with the unrestricted 
Hartree-Fock (UHF) mean-field method, we study the full description of the magnetic orders in Eu$_2$Ir$_2$O$_7$. Our numerical studies lead 
to three important conclusions listed below.  i) Besides AIAO there is an additional locally stable magnetic structure in this system, which 
is very close to AIAO in energy. The existence of this additional metastable magnetic state can perfectly explain the puzzle in the nonlinear 
magnetization measurements~\cite{liang:2017}. ii) The real ground state of Eu$_2$Ir$_2$O$_7$ is extremely sensitive to the trigonal lattice 
distortion of the pyrochlore structure and AIAO magnetic state can be stabilized only with large enough trigonal distortion. 
iii) In all these magnetic states mentioned above, in addition to the magnetic dipolar moments, we also find high-order multipolar 
moments (octupole)~\cite{santini:2009} with considerable amplitude as well. Recently, possible non-dipolar hidden order has also been implied 
from the second harmonic generation (SHG) experiments on Sr$_2$IrO$_4$~\cite{zhao:2016}. The major difference between the multipolar orders 
discussed in Sr$_2$IrO$_4$~\cite{zhao:2016} and that in the present paper is that the former one breaks the inversion symmetry. Due to the similarity 
in the local electronic structure, the results obtained in the present study may also be helpful on revealing the microscopic origin of the
hidden orders in Sr$_2$IrO$_4$.

We take the experimental lattice parameters of Eu$_2$Ir$_2$O$_7$ from Ref.~\cite{millican:2007}, that is $a=10.243 \AA$ and $x=0.3334$.
For this $x$ value, the Oxygen octahedron has a trigonal distortion (compression along the local [111] direction). We also do calculations 
for the non-distorted structure ($x=5/16$) for comparison. The DFT part of the calculations have been done by the Vienna 
Ab-initio Simulation Package (VASP)~\cite{kresse:1996}. A tight binding (TB) Hamiltonian consisting of $t_{2g}$ orbitals from four Ir atoms 
is then obtained from the non-SOC DFT calculation by the maximally localized Wannier functions method~\cite{marzari:2012, mostofi:2008}.
The $t_{2g}$ orbitals are defined with respect to the local Oxygen octahedron $XYZ$-coordinate~\cite{suppl}. An atomic SOC term is added 
to the TB Hamiltonian to account for the strong SOC of Ir atoms with its strength $\lambda$ being determined by fitting the first principle 
results. To consider the strong Coulomb interaction among Ir $t_{2g}$ orbitals, an on-site Coulomb interaction term $U$ is included. 
The total Hamiltonian can be written as,
\begin{eqnarray}
\label{eqn:ham}
H &=& \sum_{\substack{jR^{\prime}\beta\\iR\alpha}} t_{ij}^{R\alpha,R^{\prime}\beta} d_{iR\alpha}^{\dagger} d_{jR^{\prime}\beta} + \lambda \sum_{iR\alpha\beta} (\vec{\boldsymbol{l}}_{R}\cdot\vec{\boldsymbol{s}}_{R})_{\alpha\beta} d_{R\alpha}^{\dagger} d_{R\beta} \nonumber \\
 && + \frac{U}{2} \sum_{iR\alpha \beta \delta \gamma} d_{R\alpha}^{\dagger} d_{R\beta}^{\dagger} d_{R\delta} d_{R\gamma},
\end{eqnarray}
where, $i,j$ are the indices of primitive cell, $R,R^{\prime}=1,2,3,4$ are the indices of Ir sites, and $\alpha,\beta,\delta,\gamma$ are the combined 
orbital-spin indices. Under the UHF approximation, the Coulomb interaction terms are approximated as,
\begin{eqnarray}
\label{eqn:hf}
d_{R\alpha}^{\dagger} && d_{R\beta}^{\dagger} d_{R\delta} d_{R\gamma}
  \approx  \rho_{\beta\delta}^{R}  d_{R\alpha}^{\dagger} d_{R\gamma}  
        +  \rho_{\alpha\gamma}^{R} d_{R\beta}^{\dagger} d_{R\delta} 
        -  \rho_{\beta\gamma}^{R}  d_{R\alpha}^{\dagger} d_{R\delta} \nonumber \\
    & & -  \rho_{\alpha\delta}^{R} d_{R\beta}^{\dagger} d_{R\gamma} 
        -  \rho_{\beta\delta}^{R}\rho_{\alpha\gamma}^{R} + \rho_{\beta\gamma}^{R}\rho_{\alpha\delta}^{R},
\end{eqnarray}
where, $\rho_{\alpha\beta}^{R}=\Braket{d_{R\alpha}^{\dagger}d_{R\beta}}$ is the local density matrix for Ir atom at site-$R$ and is determined 
self-consistently. This TB+UHF method is numerically more stable than \textit{ab-initio} DFT+$U$ method and the total energy can converge to very high 
accuracy ($\sim 0.01$ meV)~\cite{suppl}.

The local density matrix obtained above gives the complete description of the magnetic orders in Eu$_2$Ir$_2$O$_7$. It can be decomposed 
into 36 single-particle irreducible tensor (or multipolar) operators $O_{K_{1}K_{2}}^{KM}$ defined in the spin-orbital space~\cite{santini:2009,blum:2012},
\begin{eqnarray}
\rho^{R} = \sum_{K_{1}K_{2}KM} C_{K_1K_2}^{KM} O_{K_1K_2}^{KM}, \\
C_{K_1K_2}^{KM} = \text{Tr}\left[\rho^{R}\left(O_{K_1K_2}^{KM}\right)^{\dagger}\right].
\end{eqnarray}
Note that all of these operators are defined with respect to the local [111] $xyz$-coordinate~\cite{suppl} for each Ir atom, i.e., $z$ is along 
local [111] direction. These 36 operators serve as the complete set of the 
possible order parameters (OPs) and the details of their definition can be found in the Supplementary Material (SM)~\cite{suppl}.

\begin{figure}
\includegraphics[width=0.45\textwidth]{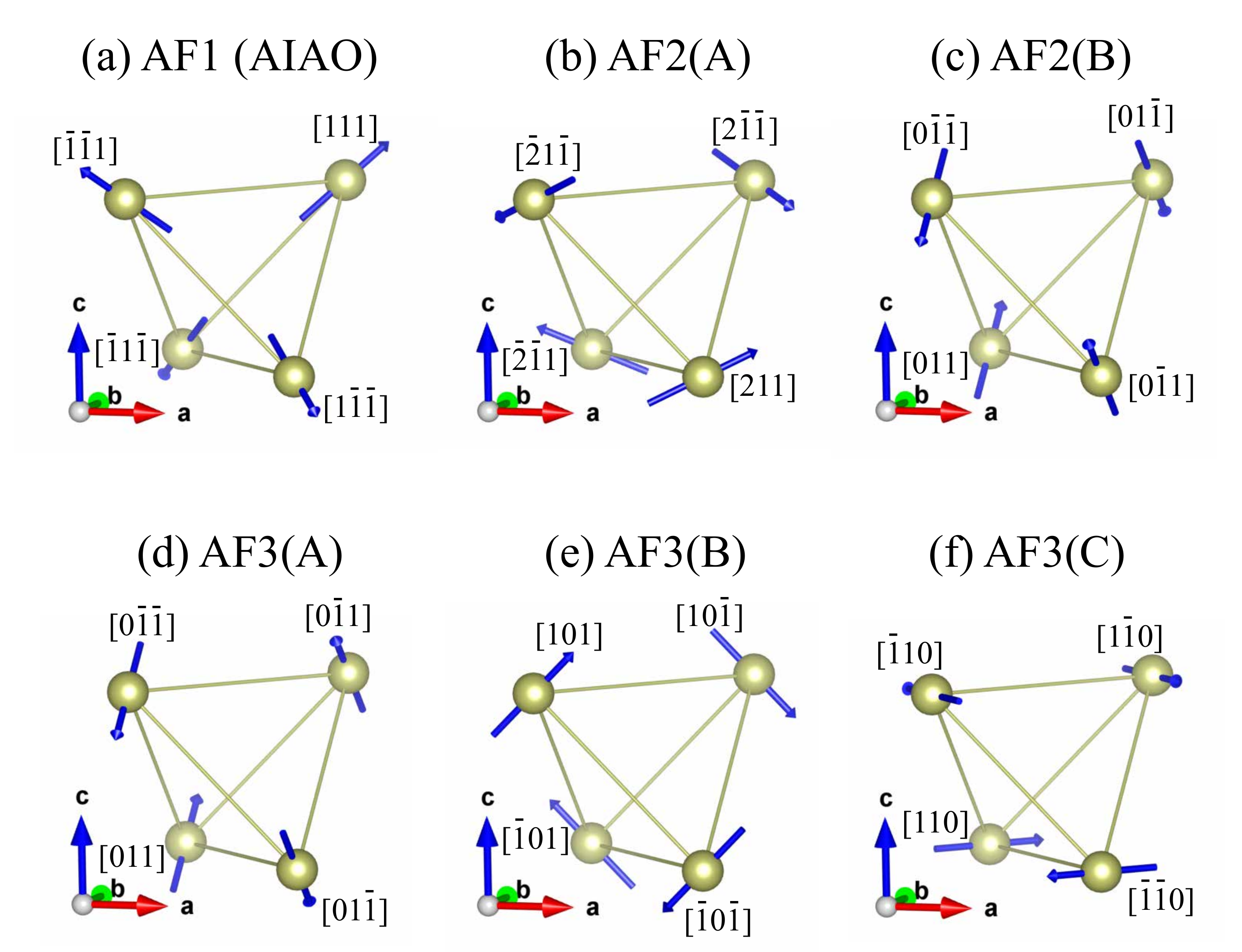}
\caption{(Color online). (a) AF1, the AIAO anti-ferromagnetic configuration, all the magnetic moments point to (against) the center of the tetrahedron.
                         (b,c) AF2, another two-fold degenerate anti-ferromagnetic configurations. 
                         (d,e,f) AF3, another three-fold degenerate anti-ferromagnetic configurations.}
\label{fig:afm}
\end{figure}

\begin{figure*}
\includegraphics[width=0.45\textwidth]{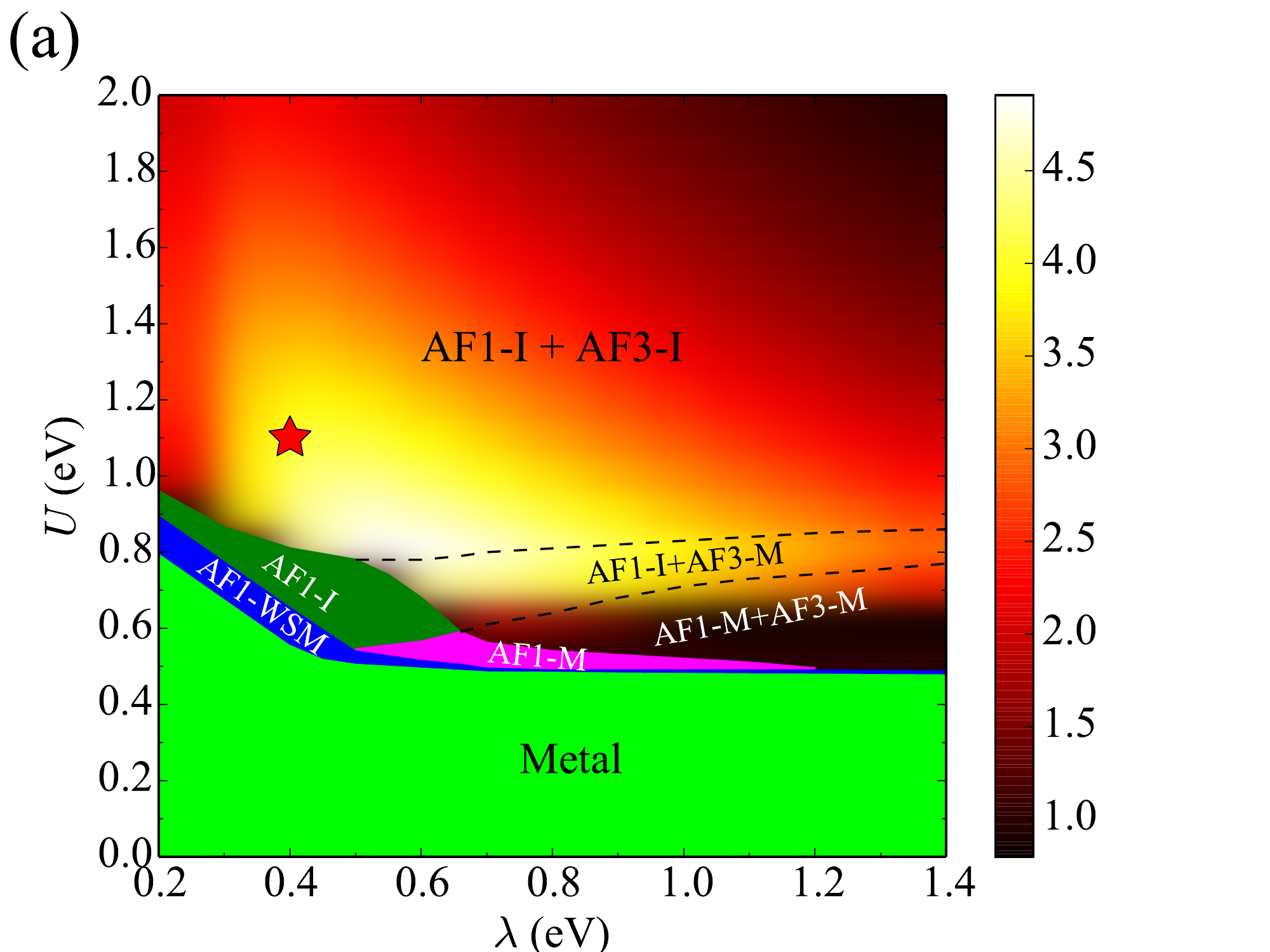}
\includegraphics[width=0.45\textwidth]{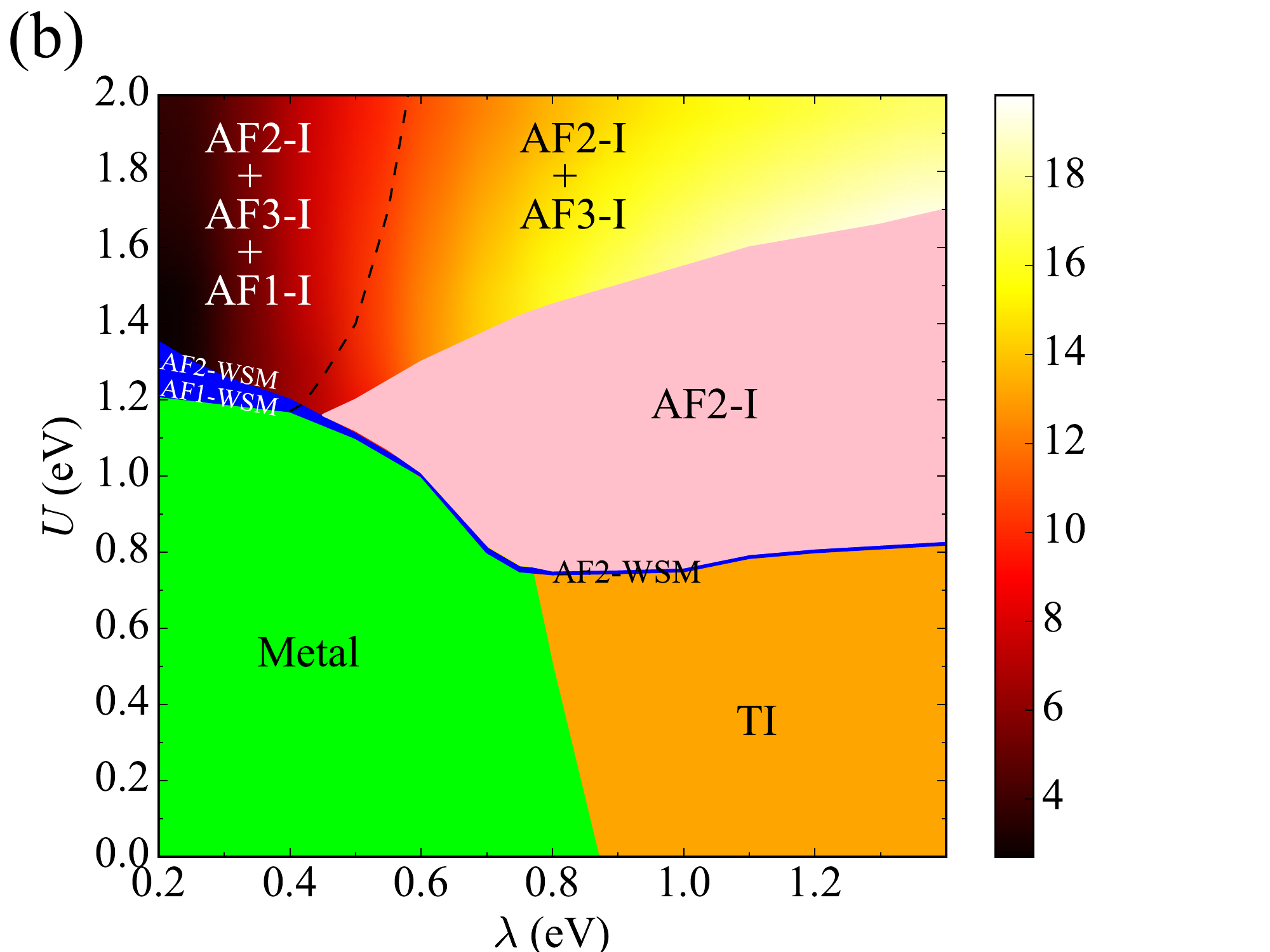}
\caption{(Color online). The phase diagram as a function of Coulomb interaction $U$ and SOC $\lambda$. 
                         (a) For the distorted structure, the ground magnetic state is always AF1. AF3 can coexist with AF1 in very large region. 
                         The colormap is used to label their total energy difference: $E_{\text{tot}}(\text{AF3})-E_{\text{tot}}(\text{AF1})$ (with units of meV).
                         The red star indicates the reasonable parameters $U=1.1$ eV and $\lambda=0.4$ eV for real material of Eu$_2$Ir$_2$O$_7$.
                         (b) For the non-distorted structure, the ground magnetic state is always AF2. AF3 and AF1 can coexist with AF2 in 
                         some regions. The colormap is used to label the total energy difference between AF3 and AF2: $E_{\text{tot}}(\text{AF3})-E_{\text{tot}}(\text{AF2})$ 
                         (with units of meV). Note that ``AF" means anti-ferromagnetic, ``I" means insulator, ``M" means metal, ``WSM" means Weyl semimetal, and ``TI" 
                         means topological insulator.}
\label{fig:phase}
\end{figure*}

\begin{figure}
\includegraphics[width=0.45\textwidth]{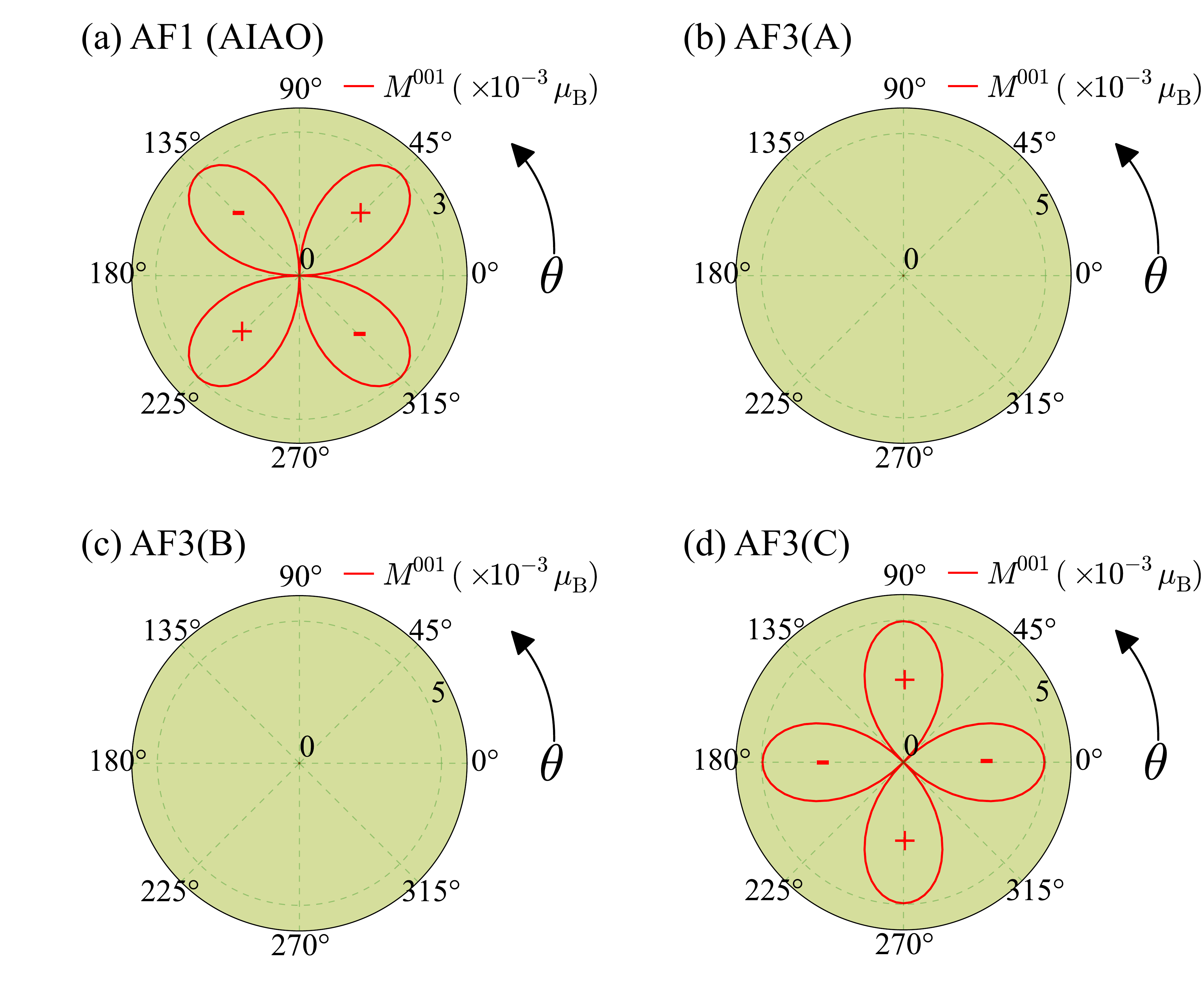}
\caption{(Color online). The net magnetic moments along the [001] direction $M^{001}$ as a function of the rotation angle $\theta$ at $U=1.1$ eV 
                         and $\lambda=0.4$ eV for (a) AF1 and (b,c,d) AF3.}
\label{fig:mag001}
\end{figure}

\begin{figure*}
\includegraphics[width=0.45\textwidth]{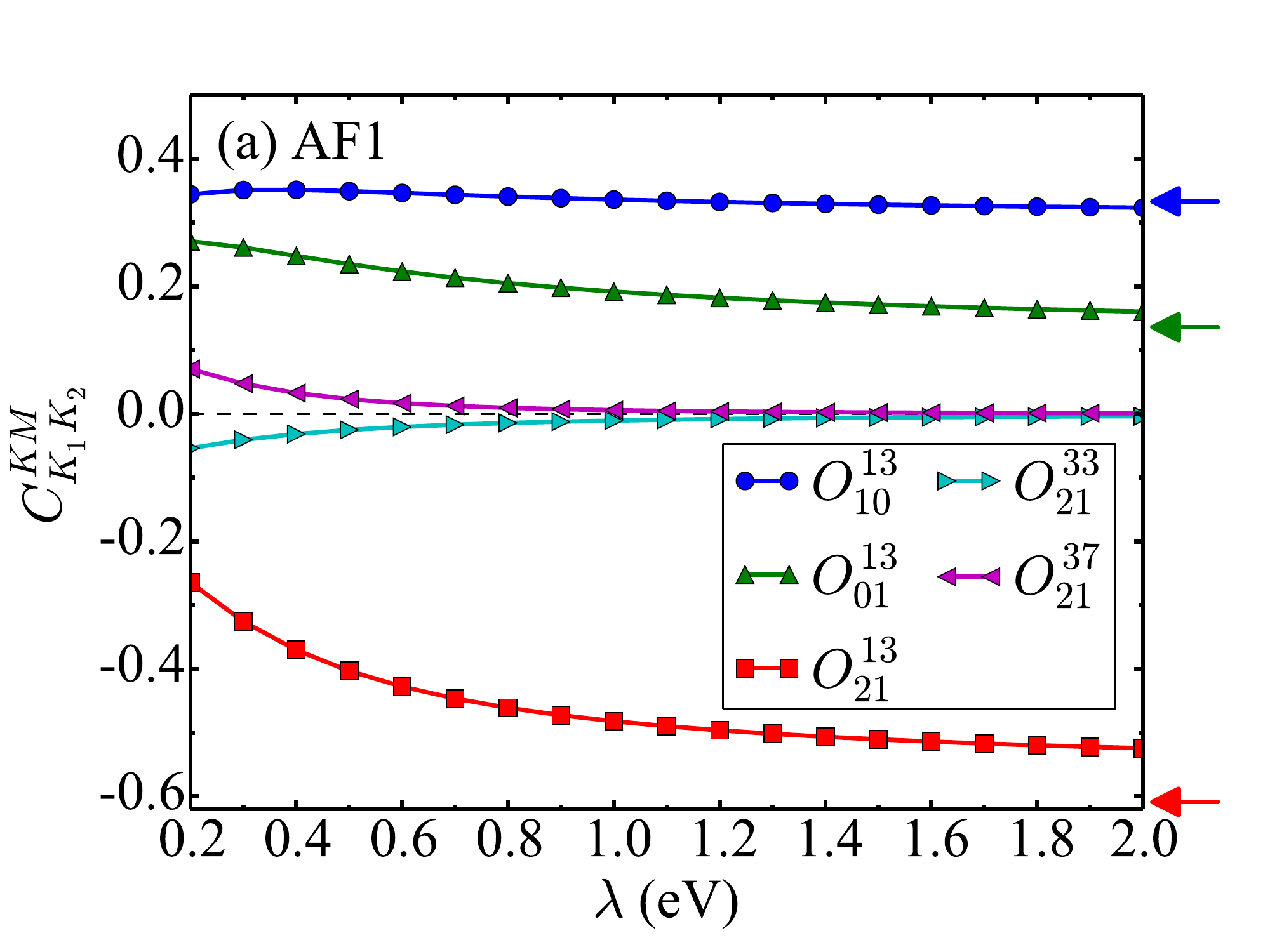}
\includegraphics[width=0.45\textwidth]{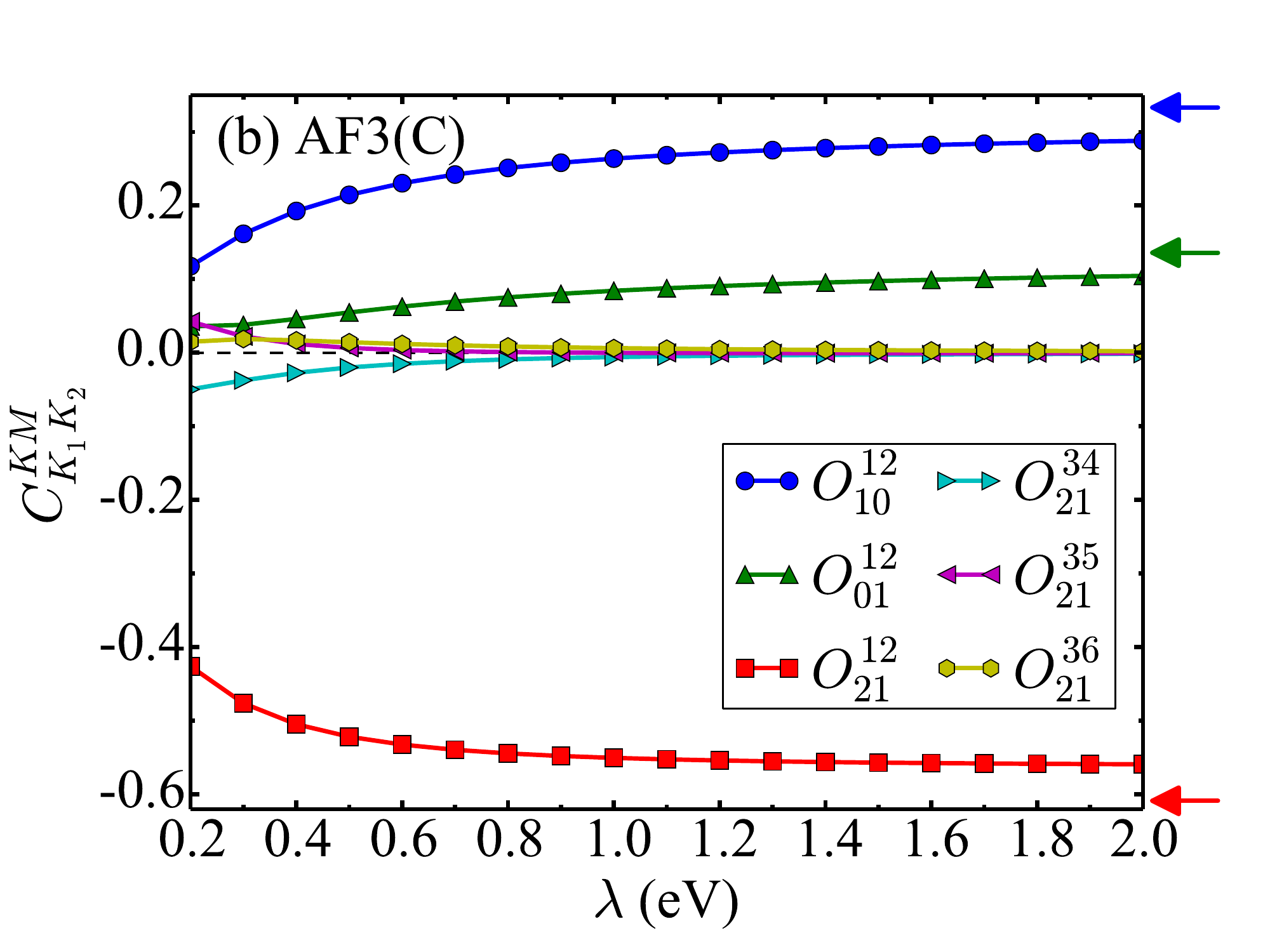}
\caption{(Color online). The weights of OPs as a function of $\lambda$ in the experimental structure with distortion at $U=1.1$ eV 
         for (a) AF1 and (b) AF3(C). The arrows indicate the expectation values of orbital (blue), spin (green) and orbital-spin (red) coupled dipoles 
         in the atomic $j_{\text{eff}}=1/2$ state, respectively. }
\label{fig:multipole}
\end{figure*}

The phase diagrams with respect to the strength of SOC $\lambda$ and Coulomb interaction $U$ for both the distorted and 
non-distorted structures are obtained and shown in Fig.~\ref{fig:phase}. With SOC, the $t_{2g}$ bands will be split into 
$j_{\text{eff}}=1/2$ and $j_{\text{eff}}=3/2$ subbands, and the $j_{\text{eff}}=1/2$ subbands are half-filled. For small $U$, the ground state 
is paramagnetic, which is unstable against magnetic oder when $U$ reaches to some critical value. Assume that the magnetic unit cell doesn't enlarge, 
all the possible magnetic structures in Eu$_2$Ir$_2$O$_7$ can be classified by finding the magnetic co-representation for the tetrahedron group,
which leads to $c\Gamma_{\text{mag}}=1c\Gamma_{3+}+1c\Gamma_{5+}+1c\Gamma_{7+}+2c\Gamma_{9+}$~\cite{wills:2006}. 
We have tried all these possible magnetic configurations in our calculations to determine the most stable magnetic order. 

Fig.~\ref{fig:phase}(a) is the phase diagram for the experimental structure of Eu$_2$Ir$_2$O$_7$, which contains finite trigonal distortion. 
The typical band structures of each phase have been plotted in the SM~\cite{suppl}. We find that the one-dimensional 
$c\Gamma_{3+}$ representation with an AIAO type anti-ferromagnetic configuration (AF1 in Fig.~\ref{fig:afm}(a)) is always the ground magnetic state, 
which is quite consistent with the previous studies~\cite{sagayama:2013,takatsu:2014,disseler:2014}. Similar to the results obtained 
in Ref.~\cite{wan:2011,ybkim:2012}, under the AF1 order, a Weyl semimetal phase can be found in a narrow region of the phase digram. With the increment 
of $U$, the Weyl semimetal phase disappears quickly leading to a semimetal to insulator transition. Besides $c\Gamma_{3+}$, we find that the 
three-dimensional $c\Gamma_{7+}$ representation (AF3 in Fig.~\ref{fig:afm}(d,e,f)), where the local moments are perpendicular to the local $[111]$ directions, 
is also stable in very large parameter region. However, its total energy per Ir atom is always a few meV higher than that of AF1 and a colormap 
is used to label their total energy difference: $E_{\text{tot}}(\text{AF3})-E_{\text{tot}}(\text{AF1})$ (with units of meV) in Fig.~\ref{fig:phase}. 
As we can see, their total energy are very close. We choose the reasonable parameters for Eu$_2$Ir$_2$O$_7$ based on the size of the 
band gap reported in previous studies~\cite{ueda:2016, hongbin:2017}, where they reported an optical gap of about 0.2 eV~\cite{ueda:2016} and a band 
gap of about 0.3 eV in their LDA+DMFT calculation~\cite{hongbin:2017}. Considering the fact that HF usually overestimates the band gap, here we choose 
a value $U=1.1$ which will induce a little larger band gap of about 0.4 eV~\cite{suppl}. A phase diagram with Hund's coupling $J_{\text{H}}$ at $J_{\text{H}}/U=0.2$ 
is also calculated and plotted in SM~\cite{suppl}. It turns out that Hund's coupling doesn't change the overall phase diagram because there is only 
one hole per Ir site in $t_{2g}$ subspace.

To study the possible nonlinear magnetization discussed in Ref.~\cite{arima:2013} and reported in Ref.~\cite{liang:2017}, we apply an external magnetic 
field $\vec{H}$ in the [001] plane and continuously rotate it by $2\pi$, and then calculate the net magnetic moments along [001] direction $M^{001}$ induced 
by the transverse magnetic field. $M^{001}$ as a function of the rotation angle $\theta$ are plotted in Fig.~\ref{fig:mag001} for AF1 and AF3 phase at $U=1.1$ eV 
and $\lambda=0.4$ eV. We only plot $M^{001}$ for one of the two time-reversal partners, and $M^{001}$ for the other partner will have the same magnitude 
but with opposite sign. Our numerical results are consistent with the experimental results, where the nonlinear magnetization pattern was rotated by 45 degree
under the field cooling process~\cite{liang:2017}. The calculated $M^{001}$ shows a $d_{xy}$ pattern for AF1, which is corresponding to the 
results in Fig.~2(a,b) in Ref.~\cite{liang:2017} where the field $\boldsymbol{H_{\text{fc}}}$ is along $[\bar{1}\bar{1}0]$ direction. While $M^{001}$ shows a 
$d_{x^2-y^2}$ pattern for AF3(C), which is corresponding to the results in Fig.~2(c,d) in Ref.~\cite{liang:2017} where the field $\boldsymbol{H_{\text{fc}}}$ 
is along $[\bar{1}10]$ direction. Note that the measured nonlinear magnetization in Ref.~\cite{liang:2017} show distorted $d$-waves which may be caused 
by an additional order which already exists at 300 K. $M^{001}$ for AF3(A) and AF3(B) configurations are zero due to the symmetry reason. The occurrence of 
additional metastable magnetic phase AF3 can explain the observed $\pi/4$ phase shift of the magnetization pattern with the assumption that the field cooling 
processes may stabilize AF3. 

Fig.~\ref{fig:phase}(b) is the phase diagram for the ideal pyrochlore structure without any distortion. We find that the two-dimensional $c\Gamma_{5+}$ 
representation (AF2 in Fig.~\ref{fig:afm}(b,c)) is always the ground state, which is quite different with the situation in the distorted structure. 
AF3 and AF1 are both locally stable metastable states here and can also coexist with AF2 in some parameter region. The total energy of all the three phases 
satisfies $E_{\text{tot}}(\text{AF2})<E_{\text{tot}}(\text{AF3})<E_{\text{tot}}(\text{AF1})$. The colormap is used to label the total energy difference 
between AF3 and AF2: $E_{\text{tot}}(\text{AF3})-E_{\text{tot}}(\text{AF2})$ (with units of meV). These results indicate that a large enough trigonal 
distortion~\cite{uematsu:2015} may be crucial for stabilizing the AIAO state. For most of the pyrochlore iridates, the trigonal distortion 
is indeed large enough, which implies that the AIAO is likely to be the ground state. 

In the anti-ferromagnetic phase of the distorted structure, we expand the local density matrix $\rho$ to measure the weights of the OPs.
We plot their weights as a function of $\lambda$ for AF1 and AF3(C) phases at $U=1.1$ eV in Fig.~\ref{fig:multipole}. The new finding of 
our calculations is that, besides large dipoles, there also exist considerable higher-rank octupolar moments. 

In AF1 phase, the dipoles are $O_{10}^{13}=l_{z}/2$ and $O_{01}^{13}=\sqrt{2/3}s_{z}$. Besides these dipoles, there is another spin-orbital coupled dipole 
$O_{21}^{1}$~\cite{carra:1993,luo:1993} with $O_{21}^{11}$, $O_{21}^{12}$ and $O_{21}^{13}$ being the $x$-,$y$- and $z$-components (Eqn.~S45-S47 in SM~\cite{suppl}). 
In AF1 phase, the nonzero component is $O_{21}^{13}$. In AF3(C) phase, the non-zero components of dipoles are $O_{10}^{12}=l_{y}/2$, $O_{01}^{12}=\sqrt{2/3}s_{y}$ 
and $O_{21}^{12}$. The arrows on the right side of Fig.~\ref{fig:multipole}(a,b) mark the expectation values of the components of $O_{10}^{1}$ (blue), $O_{01}^{1}$ (green) 
and $O_{21}^{1}$ (red) in the ideal atomic $j_{\text{eff}}=1/2$ state (large SOC limit). At small SOC, their expectation values deviate quite far away from 
their atomic limits, and approach to the atomic limits with the increment of SOC. At $U=1.1$ eV and $\lambda=0.4$ eV, the calculated ratio $\Braket{L_{z}}/\Braket{S_{z}}$ 
is about 2.3 for AF1 and the ratio $\Braket{L_{y}}/\Braket{S_{y}}$ is about 6.8 for AF3(c), which deviates quite far away from the value 4 in the 
atomic $j_{\text{eff}}=1/2$ states~\cite{laguna:2010,fujiyama:2014}. These results indicate that the mixing of $j_{\text{eff}}=1/2$ and $j_{\text{eff}}=3/2$ 
states is indeed significant in Eu$_2$Ir$_2$O$_7$ and the $j_{\text{eff}}=1/2$ single orbital picture is not applicable here.

The octupoles are defined as the components of a rank-3 irreducible tensor. We find two components $O_{21}^{33}$, $O_{21}^{37}$ (Eqn.~S55 and Eqn.~S59 in SM~\cite{suppl}) 
with nonzero weights in AF1 phase, and three components $O_{21}^{34}$, $O_{21}^{35}$, $O_{21}^{36}$ (Eqn.~S56-S58 in SM~\cite{suppl}) in AF3(C) phase.
As shown in Fig.~\ref{fig:multipole}, for both AF1 and AF3(C) phases, with the realistic SOC strength ($\lambda = 0.4$ eV) the octupole weight can be 
comparable with the dipoles and cannot be ignored. With the increment of SOC, the octupole weights will vanish gradually and the magnetic moments
of Iridium ions can be described by the dipole moments only formed within the $j_{\text{eff}}=1/2$ subspace in the strong SOC limit.
We would emphasize that the mixing of $j_{\text{eff}}=1/2$ and $j_{\text{eff}}=3/2$ subspaces to some extent is the prerequisite for the 
occurrence of octupoles because $j_{\text{eff}}=1/2$ subspace alone can only induce multipolar moments up to rank-1. We also note that the size ratio of 
octupoles and dipoles in AF3 phase is a little larger than that in AF1 phase, implying that the effective spin-orbit coupling $\Braket{\boldsymbol{L}\cdot\boldsymbol{S}}$ 
would be smaller in AF3 than in AF1. This change may be seen in the XAS Ir $L_{2}/L_{3}$ branching ratio~\cite{laguna:2010,clancy:2012}. The occurrence of 
these octupolar moments may bring some interesting physical consequence, which needs further study.

In summary, we have found stable AIAO magnetic ground state in Eu$_2$Ir$_2$O$_7$ only when the trigonal lattice distortion is fully considered. 
Besides AIAO, a metastable magnetic phase AF3 with very close energy to AIAO is also found. The appearance of AF3 phase can explain the nonlinear 
magnetization behavior in Eu$_2$Ir$_2$O$_7$ under field cooling. In the magnetic phase, besides large dipoles, we also find high-order multipolar 
octupoles with considerable amplitude. These results of Eu$_2$Ir$_2$O$_7$ serve as an example and can be used to explain the magnetic properties 
for other pyrochlore iridates. 

Y.L.W. acknowledges the helpful discussions with Dr. Jianzhou Zhao and Rui Yu about the calculations. 
X.D. acknowledges the helpful discussions with Professor Nai Phuan Ong, Dr. Tian Liang and Professor Hiroshi Shinaoka.
This work was supported by the NSF of China and by the 973 program of China (No. 2011CBA00108 and No. 2013CB921700). 

\bibliography{Eu2Ir2O7}

\end{document}